# Title: A Voltage Calibration Chain for Meters Used in Measurements of EV Inductive Power Charging

Authors: M. Zucca, U. Pogliano, M. Modarres, D. Giordano, G. Crotti, D. Serazio

**Acknowledgement:** The results presented in this paper are developed in the framework of the 16ENG08 MICEV Project. The latter received funding from the EMPIR programme co-financed by the Participating States and from the European Union's Horizon 2020 research and innovation programme.

# A voltage calibration chain for meters used in measurements of EV inductive power charging

M. Zucca, U. Pogliano, M. Modarres, D. Giordano, G. Crotti, D. Serazio

Istituto Nazionale di Ricerca Metrologica, INRIM, Strada delle Cacce 91, 10125, Torino, Italy
m.zucca@inrim.it

***Abstract* — The inductive charging of electric vehicles requires specific measurement and calibration systems. In fact, the measurement of power on board involves DC signals, which are superimposed to a significant AC ripple up to or over 150 kHz, depending on the type of charging system. A calibration method that makes use of a phantom power, based on two independent but synchronized circuits, is considered, simulating the charging voltage and current. This paper describes in detail a solution in the realization of the voltage calibration chain, based on the use of a DC voltage calibrator, an injector and a voltage divider.**

## I. Introduction

Accurate power and efficiency measurements are very important for electric automotive application and design. High-accuracy voltage measurements become critical when frequency bandwidth and voltage magnitude increase since many commercially available probes show low accuracy, especially at increasing frequency [1]. The inductive power transfer (IPT) for vehicles charging is an electrical automotive application of growing interest. IPT occurs between a coil located at about ground level and a coil placed on board, through the coupling of a magnetic field. The coils are both connected to a resonant circuit that resonates at the same frequency, which can vary from 20 kHz to over 100 kHz, being a typical resonance frequency equal to 85 kHz [2]. An AC-DC converter then rectifies the coil signal induced on board; the voltage to be measured at the battery being charged is a DC signal with the superposition of a periodic ripple. A laboratory facility suited to perform the calibration of sensors for voltage measurement is described in the following. The laboratory facility includes a section for generating the signal, with DC voltages that can reach 1000 volts with an AC ripple up to 15% of the total voltage (less than 150 V peak). The facility also includes the measurement section, which comprises a reference sensor. Given the voltage magnitude and the frequency bandwidth, a divider designed for the purpose is the most suitable solution. A target relative uncertainty of $0.5 \cdot 10^{-3}$ is expected for the final setup.

## II. Performances of a commercial voltage probe

As a first step, the frequency performances of a commercial differential active voltage probe used for measurement in EV

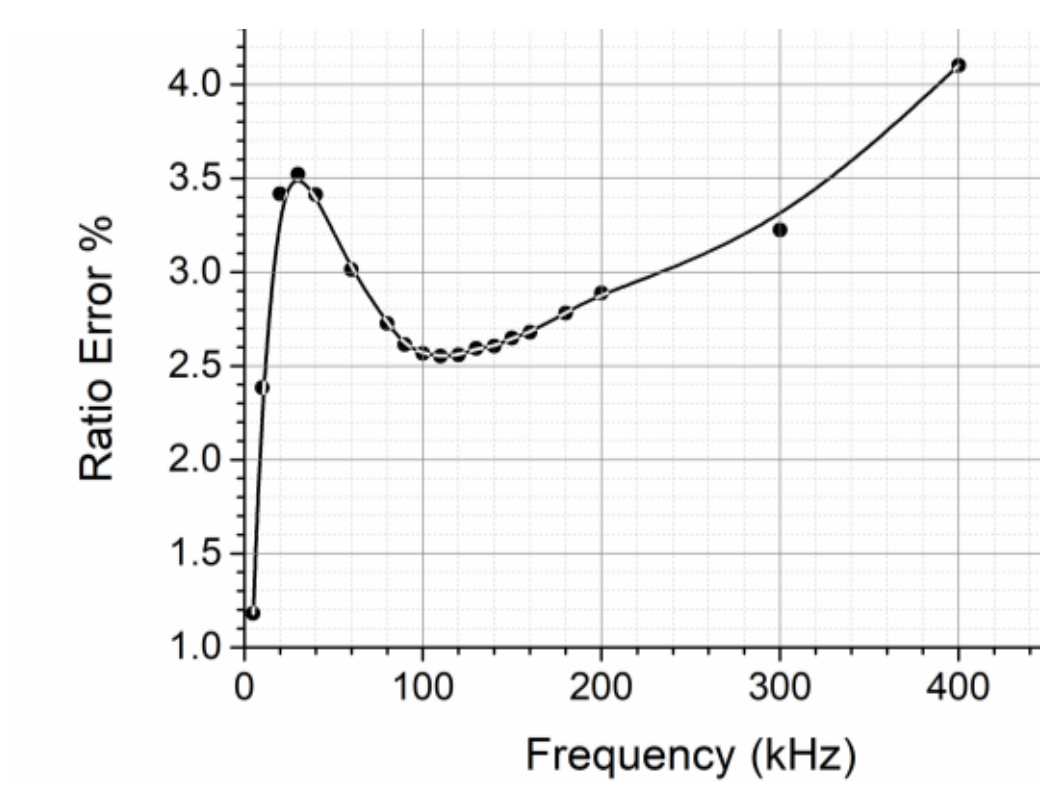

Fig. 1 – Percentage ratio error of the differential probe under test

technology development are investigated. The rated specifications are: bandwidth 20 MHz, differential voltage 1400 $V_{pk}$, best low-frequency accuracy 1% of reading. The ratio error is obtained by comparing the applied voltage measured by a Keysight 3458A multimeter with the probe output measured by an oscilloscope. Fig. 1 shows the measured ratio error where an AC ripple having magnitude 282 V peak to peak has been applied to the probe. It can be noted how the rated accuracy is found only at power frequency, while in the frequency range of interest (20 kHz – 150 kHz) the error varies from 2.5 % to 3.5 %. Due to these figures, a dedicated calibration system for such probes under operating conditions is needed.

## III. Calibration circuit

### *A. Supply Section*

Fig. 2 shows the scheme of the voltage calibration circuit. It consists of a voltage calibrator (Fluke 5500), which provides a stable DC voltage. Series connected to the calibrator, a voltage injector designed for the purpose is included in the circuit, supplied by an AC amplifier (NF Model HSA4052) driven by a signal generator. The voltage injector is a wideband transformer, which is obtained by a proper core and two windings. Two windings with the same number of turns have been realized. In this way, the voltage transformer can be easily characterized by simply measuring the difference between the voltages at the two windings when they are connected in opposition.

The core has been made of nanocrystalline material

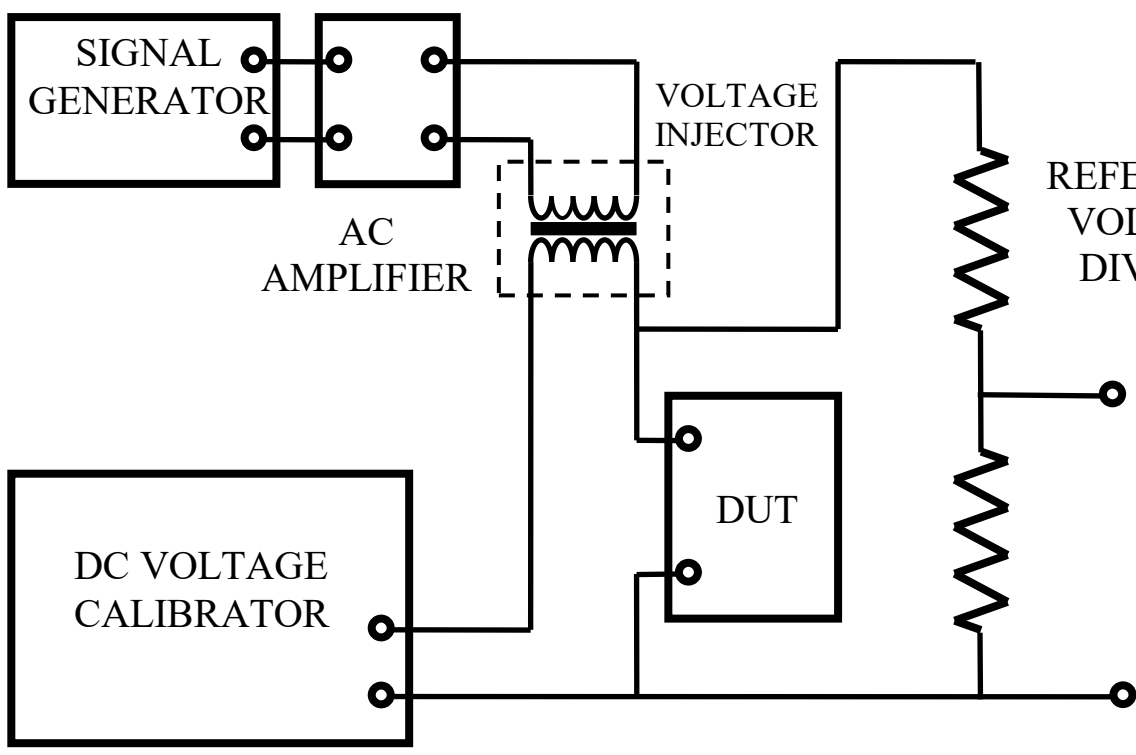

Fig. 2 - Scheme of the voltage calibration circuit

(Vitroperm 500), whose relative magnetic permeability varies almost linearly in the logarithmic scale from $\mu_r \approx 80000$ at 20 kHz up to $\mu_r \approx 8000$ at 1 MHz. Such a linear behaviour allowed an easy estimation of the winding inductances versus frequency. The saturation magnetization of Vitroperm is 1.2 T. Having established the geometry of the core (W424 from Vacuumschmelze), the maximum operating voltage (200 V peak) and the saturation induction, the minimum number of turns has been determined equal to 20 for both windings.

Subsequently, the transformer frequency performances were estimated using the T-circuit as a representation, and Mathcad ™ as a tool for simulations. The network of stray capacitances between the winding turns has been taken into account and the injector scheme and its final parameters are reported in Fig. 3 and Table 1, respectively. Such design guarantees a flat response of the injector up to more than 150 kHz, covering the bandwidth for the considered application.

A first measurement test with the two windings connected in opposition, being the frequency equal to 85 kHz, has shown a ratio error of about 75µV/V and a phase difference of 37 µrad.

### *B. Measurement section*

The measurement section consists of a reference voltage divider and an acquisition system. The divider can be a resistive or resistive-capacitive divider. Prototypes of both types of divider have been designed and realized based on the experience [3-5] and their response is under evaluation by a calibration procedure (see for example [5-6]) based, for this specific purpose, on the stability of the DC source and the accuracy of both the DC voltage and the injector ratio. The voltage section of the device under test is connected in parallel to the input of the divider and calibrated, taking into account, via software, the divider and the acquisition system errors.

| Parameter | Value | Parameter | Value |
|---|---|---|---|
| core | VAC W424 | $R_{s2}$ | 0. 153 Ω |
| turns | 20 in two layers | $L_{s1}$ | 0.795 µH |
| $L_p$ (20 kHz) | 33.7 mH | $L_{s2}$ | 1.26 µH |
| $L_p$ (1 MHz) | 3.37 mH | $C_{p1}$ | 36.65 pF |
| $Res_p$ | 10.11 kΩ | $C_{p2}$ | 39.32 pF |
| $R_{s1}$ | 0.118 Ω | | |

**Table 1**: Computed parameters of the injector

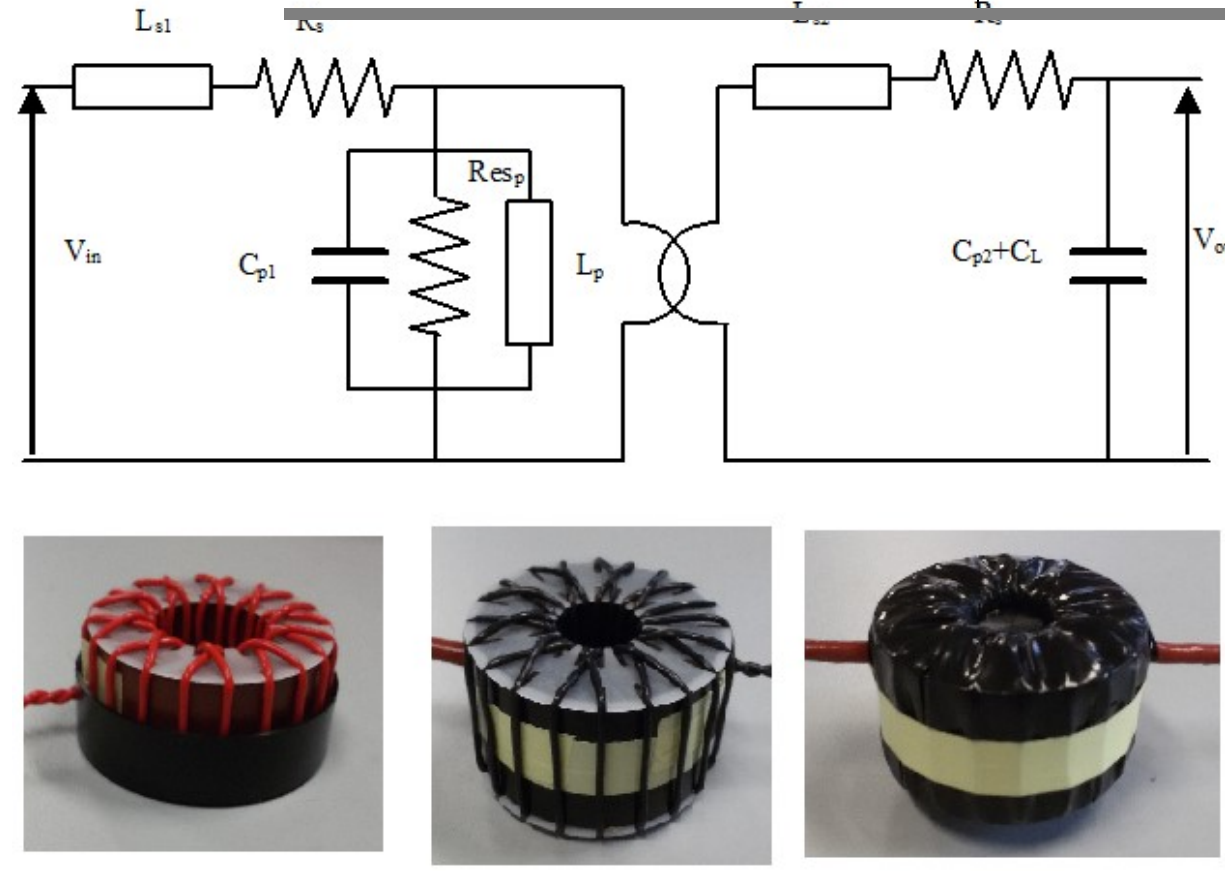

Fig. 3 – Circuital representation of the voltage injector and images of its realization

## VI. CONCLUSION

A system for accurate characterizations of instruments used in voltage and power measurements, used on board EV's equipped with IPT systems, has been designed and is currently under tuning. The system is built by a combination of a high stable and accurate sources, capable of generating DC signals with a superimposed ripple up to 150 kHz and reference voltage dividers. After the completion of the calibration of the dividers in ratio and phase, it will be the INRIM reference for the measurements in automotive electric applications.

## ACKNOWLEDGEMENT

The results here presented are developed in the framework of the European EMPIR 16ENG08 MICEV Project.